\documentclass[12pt]{article}
\usepackage{epsfig}

\newcommand\pubnumber{}
\newcommand\pubdate{May 2001}
\newcommand\hepnumber{hep-ph/0106020}

\def\csumb{$^a$Department of Physics and Astronomy, UCLA,\\
        \em Los Angeles, California 90095-1547, USA\\
  \em $^b$Randall Laboratory of Physics, University of Michigan,\\
      \em Ann Arbor, Michigan 48109-1120, USA}
\def\support{\footnote{Work supported by the US Department of Energy.}} 

\def\Title#1{\begin{center} {\Large\bf #1 } \end{center}}
\def\Author#1{\begin{center}{ \sc #1} \end{center}}
\def\Address#1{\begin{center}{ \it #1} \end{center}}

\newcommand\pubblock{\rightline{\begin{tabular}{l} \pubnumber\\
         \pubdate\\ \hepnumber \end{tabular}}}
\newenvironment{Abstract}{\begin{quotation}  }{\end{quotation}}
\newenvironment{Presented}{\begin{quotation} \begin{center} 
             Presented at the\end{center}
      \begin{center}\begin{large}}{\end{large}\end{center} \end{quotation}}

\makeatletter
\def\section{\@startsection{section}{0}{\z@}{5.5ex plus .5ex minus
 1.5ex}{2.3ex plus .2ex}{\large\bf}}
\def\subsection{\@startsection{subsection}{1}{\z@}{3.5ex plus .5ex minus
 1.5ex}{1.3ex plus .2ex}{\normalsize\bf}}
\def\subsubsection{\@startsection{subsubsection}{2}{\z@}{-3.5ex plus
-1ex minus  -.2ex}{2.3ex plus .2ex}{\normalsize\sl}}

\renewcommand{\@makecaption}[2]{%
   \vskip 10pt
   \setbox\@tempboxa\hbox{\small #1: #2}
   \ifdim \wd\@tempboxa >\hsize     
       \small #1: #2\par          
     \else                        
       \hbox to\hsize{\hfil\box\@tempboxa\hfil}
   \fi}

 \def\citenum#1{{\def\@cite##1##2{##1}\cite{#1}}}
 
\newcount\@tempcntc
\def\@citex[#1]#2{\if@filesw\immediate\write\@auxout{\string\citation{#2}}\fi
  \@tempcnta\z@\@tempcntb\m@ne\def\@citea{}\@cite{\@for\@citeb:=#2\do
    {\@ifundefined
       {b@\@citeb}{\@citeo\@tempcntb\m@ne\@citea\def\@citea{,}{\bf ?}\@warning
       {Citation `\@citeb' on page \thepage \space undefined}}%
    {\setbox\z@\hbox{\global\@tempcntc0\csname b@\@citeb\endcsname\relax}%
     \ifnum\@tempcntc=\z@ \@citeo\@tempcntb\m@ne
       \@citea\def\@citea{,}\hbox{\csname b@\@citeb\endcsname}%
     \else
      \advance\@tempcntb\@ne
      \ifnum\@tempcntb=\@tempcntc
      \else\advance\@tempcntb\m@ne\@citeo
      \@tempcnta\@tempcntc\@tempcntb\@tempcntc\fi\fi}}\@citeo}{#1}}
\def\@citeo{\ifnum\@tempcnta>\@tempcntb\else\@citea\def\@citea{,}%
  \ifnum\@tempcnta=\@tempcntb\the\@tempcnta\else
  {\advance\@tempcnta\@ne\ifnum\@tempcnta=\@tempcntb \else\def\@citea{--}\fi
    \advance\@tempcnta\m@ne\the\@tempcnta\@citea\the\@tempcntb}\fi\fi}
\makeatother

\def\beq{\begin{equation}}
\def\eeq#1{\label{#1}\end{equation}}
\def\eeqn{\end{equation}}

\newenvironment{Eqnarray}%
   {\arraycolsep 0.14em\begin{eqnarray}}{\end{eqnarray}}
\def\beqa{\begin{Eqnarray}}
\def\eeqa#1{\label{#1}\end{Eqnarray}}
\def\eeqan{\end{Eqnarray}}

\let\bar=\overbar

\def\Dslash{\not{\hbox{\kern-4pt $D$}}}
\def\dslash{\not{\hbox{\kern-2pt $\del$}}}

\def\msb{{\bar{\ssstyle M \kern -1pt S}}}

\def\lsim{\mathrel{\raise.3ex\hbox{$<$\kern-.75em\lower1ex\hbox{$\sim$}}}}
\def\gsim{\mathrel{\raise.3ex\hbox{$>$\kern-.75em\lower1ex\hbox{$\sim$}}}}

\begin{document}
\begin{titlepage}
\pubblock

\vfill
\def\thefootnote{\fnsymbol{footnote}}
\Title{Electroweak radiative corrections: \\[5pt] 
      Towards a full two-loop analysis}
\vfill
\Author{Adrian Ghinculov$^a$\support and York-Peng Yao$^b$\support}
\Address{\csumb}
\vfill
\begin{Abstract}
In calculating electroweak radiative corrections at two-loop level,
one encounters Feynman graphs with several different masses on the internal
propagators and on the external legs, which lead to complicated scalar 
functions.
We describe a general analytic-numerical reduction scheme for evaluating 
any two-loop diagrams with general kinematics and general 
renormalizable interactions, whereby ten basic functions form a
complete set after tensor reduction. 
We illustrate this scheme by applying it to two- and three-point functions. 
We discuss the treatment of infrared singularities within this numerical 
approach.
\end{Abstract}
\vfill
\begin{Presented}
5th International Symposium on Radiative Corrections \\ 
(RADCOR--2000) \\[4pt]
Carmel CA, USA, 11--15 September, 2000
\end{Presented}
\vfill
\end{titlepage}
\def\thefootnote{\arabic{footnote}}
\setcounter{footnote}{0}

Because of the level of experimental precision attained in measuring the
electroweak parameters at LEP, SLC, and Tevatron, a full two-loop analysis
is desirable. This will be even more necessary in view of the precision 
envisioned at future colliders such as the LHC, NLC, and the GigaZ. 

While several two-loop quantities are already included in 
standard electroweak fitting programs such as ZFITTER \cite{zfitter}, 
they are often 
obtained within certain approximations where one can neglect certain masses 
or can perform a mass expansion of Feynman graphs. These approximation 
techniques, while ingenious, were used because the exact Feynman integrals 
typically lead to complicated scalar functions which often cannont be 
evaluated analytically, in a closed form, in terms of usual special functions.
During the past decade, the existing work on massless \cite{qcd}
and massive \cite{weiglein:1}--\cite{topexpansion} two-loop graphs made 
it clear that for the general mass case, a certain amount of numerical work 
is unavoidable. 

Here we discuss the status of a hybrid, analytical-numerical approach to
two-loop radiative corrections with arbitrary masses. The aim is to treat 
any two-loop graph, of any topology, by using the same algorithm, 
so that the general recipe can be encoded in a computer program. 
Such an approach was developed in ref. \cite{2loopgeneral}, 
and was successfully applied to several physical processes 
\cite{onshellexp}--\cite{2loopnumerical}.

At the center of this approach is the introduction of a set of ten basic 
functions, $h_1$---$h_{10}$, defined by the following one-dimensional
integral representations:

\begin{eqnarray}
    h_1(m_1,m_2,m_3;k^2) & = &  \int_0^1 dx \,
                                \tilde{g} (x)
  \nonumber \\
    h_2(m_1,m_2,m_3;k^2) & = &  \int_0^1 dx \,
                              [ \tilde{g}   (x)
                              + \tilde{f_1} (x) ]
  \nonumber \\
    h_3(m_1,m_2,m_3;k^2) & = &  \int_0^1 dx \, 
                              [ \tilde{g}   (x)
                              + \tilde{f_1} (x) ] \, (1-x)
  \nonumber \\
    h_4(m_1,m_2,m_3;k^2) & = &  \int_0^1 dx \,
                              [ \tilde{g}   (x)
                              + \tilde{f_1} (x)
                              + \tilde{f_2} (x) ]
  \nonumber \\
    h_5(m_1,m_2,m_3;k^2) & = &  \int_0^1 dx \,
                              [ \tilde{g}   (x)
                              + \tilde{f_1} (x)
                              + \tilde{f_2} (x) ] \, (1-x)
  \nonumber \\
    h_6(m_1,m_2,m_3;k^2) & = &  \int_0^1 dx \,
                              [ \tilde{g}   (x)
                              + \tilde{f_1} (x)
                              + \tilde{f_2} (x) ] \, (1-x)^2
  \nonumber \\
    h_7(m_1,m_2,m_3;k^2) & = &  \int_0^1 dx \,
                              [ \tilde{g}   (x)
                              + \tilde{f_1} (x)
                              + \tilde{f_2} (x)
                              + \tilde{f_3} (x) ]
  \nonumber \\
    h_8(m_1,m_2,m_3;k^2) & = &  \int_0^1 dx \,
                              [ \tilde{g}   (x)
                              + \tilde{f_1} (x)
                              + \tilde{f_2} (x)
                              + \tilde{f_3} (x) ] \, (1-x)
  \nonumber \\
    h_9(m_1,m_2,m_3;k^2) & = &  \int_0^1 dx \,
                              [ \tilde{g}   (x)
                              + \tilde{f_1} (x)
                              + \tilde{f_2} (x)
                              + \tilde{f_3} (x) ] \, (1-x)^2
  \nonumber \\
    h_{10}(m_1,m_2,m_3;k^2) & = &  \int_0^1 dx \,
                              [ \tilde{g}   (x)
                              + \tilde{f_1} (x)
                              + \tilde{f_2} (x)
                              + \tilde{f_3} (x) ] \, (1-x)^3
   \; \; ,
\label{eq:intrepr}
\end{eqnarray}
where:

\begin{eqnarray}
  \tilde{g} (m_1,m_2,m_3;k^2;x) & = &
     Sp(\frac{1}{1-y_1}) 
   + Sp(\frac{1}{1-y_2}) 
   + y_1 \log{\frac{y_1}{y_1-1}} 
   + y_2 \log{\frac{y_2}{y_2-1}} 
  \nonumber \\
  \tilde{f_1}(m_1,m_2,m_3;k^2;x) & = &
   \frac{1}{2}
   \left[
   - \frac{1-\mu^2}{\kappa^2}
   + y_1^2 \log{\frac{y_1}{y_1-1}} 
   + y_2^2 \log{\frac{y_2}{y_2-1}} 
   \right]
  \nonumber \\
  \tilde{f_2}(m_1,m_2,m_3;k^2;x) & = &
   \frac{1}{3}
   \left[
   - \frac{2}{\kappa^2} 
   - \frac{1-\mu^2}{2 \kappa^2}
   - \left( \frac{1-\mu^2}{\kappa^2} \right)^2
   \right.
  \nonumber \\
 & &
   \; \; \; \; \; \; \; \; \; \; \; \;
   \; \; \; \; \; \; \; \; \; \; \; \;
   \; \; \; \; \; \; \; \; \; \; \; \;
   \left.
   + y_1^3 \log{\frac{y_1}{y_1-1}} 
   + y_2^3 \log{\frac{y_2}{y_2-1}} 
   \right]
  \nonumber \\
  \tilde{f_3}(m_1,m_2,m_3;k^2;x) & = &
   \frac{1}{4}
   \left[
   - \frac{4}{\kappa^2} 
   - \left( \frac{1}{3} + \frac{3}{\kappa^2}  \right) 
     \left( \frac{1-\mu^2}{\kappa^2} \right)
   - \frac{1}{2} \left( \frac{1-\mu^2}{\kappa^2} \right)^2
   - \left( \frac{1-\mu^2}{\kappa^2} \right)^3
   \right.
  \nonumber \\
 & &
   \left.
   \; \; \; \; \; \; \; \; \; \; \; \;
   \; \; \; \; \; \; \; \; \; \; \; \;
   \; \; \; \; \; \; \; \; \; \; \; \;
   + y_1^4 \log{\frac{y_1}{y_1-1}} 
   + y_2^4 \log{\frac{y_2}{y_2-1}} 
   \right]
   \; \; .
\end{eqnarray}
Here we use the following notations:

\begin{eqnarray}
y_{1,2} & = & \frac{1 + \kappa^{2} - \mu^{2}
                    \pm \sqrt{\Delta}}{2 \kappa^{2}}  \nonumber \\
\Delta  & = & (1 + \kappa^{2} - \mu^{2})^{2} 
          + 4 \kappa^{2} \mu^{2} - 4 i \kappa^{2} \eta 
      \; \; ,
\end{eqnarray}
and

\begin{eqnarray}
   \mu^{2}  & = &  \frac{a x + b (1-x)}{x (1-x)}   \nonumber \\
         a  & = &  \frac{m_{2}^{2}}{m_{1}^{2}} \, , \; \; \; \;
         b \; = \; \frac{m_{3}^{2}}{m_{1}^{2}} \, , \; \; \; \;
\kappa^{2} \; = \; \frac{    k^{2}}{m_{1}^{2}} 
      \; \; .
\end{eqnarray}

The evaluation of these ten functions is best done by numerical integration.
By using an adaptative deterministic numerical integration algorithm, these
one-dimensional integrals can be calculated fast and very precisely. 
By numerical integration, we plot these ten basic functions in figure 1 for a 
range of their kinematic variables.

\begin{figure}[t]
    \epsfxsize = 12cm
\begin{center}
    \epsffile{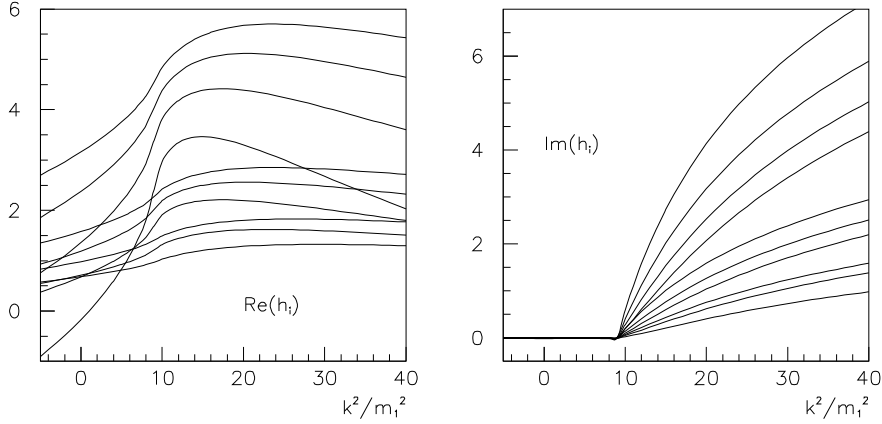}
\end{center}
\caption{Plots of the ten basic functions $h_i(m_1^2,m_2^2,m_3^2;-k^2)$ 
         as a function of the external momentum variable $k^2$. The
         plots given here are for $m_1^2=m_2^2=m_3^2=1$.}
\end{figure}

Within the method we discuss here, any two-loop diagram with arbitrary 
masses is first reduced to multi-dimensional scalar integrals involving 
these ten basic functions $h_i$. In ref. \cite{2loopgeneral} we have 
shown that this can always be done for any two-loop diagrams occuring 
in renormalizable theories. We note in passing that for the case of 
non-renormalizable theories this set of ten functions $h_i$ in general 
needs to be extendeed to include additional functions.

This reduction procedure starts by introducing Feynman parameters in the
original Feynman graph in order to relate the integral to sunset-type 
integrals. This is illustrated in figure 2.

At his point, the Feynman graph is written as a multi-dimensional integral
over tensor integrals of the following type:

\begin{equation}
   \int d^{n}p\,d^{n}q\, 
       \frac{p^{\mu_1} \ldots p^{\mu_i} q^{\mu_{i+1}} \ldots q^{\mu_j}}{
             [(p+k)^{2}+m_{1}^{2}]^{\alpha_{1}} \,
             (q^{2}+m_{2}^{2})^{\alpha_{2}} \,
             (r^{2}+m_{3}^{2})^{\alpha_{3}}
	    }
    ~~ .
\label{eq:1}
\end{equation}

\begin{figure}[t]
    \epsfxsize = 9cm
\begin{center}
    \epsffile{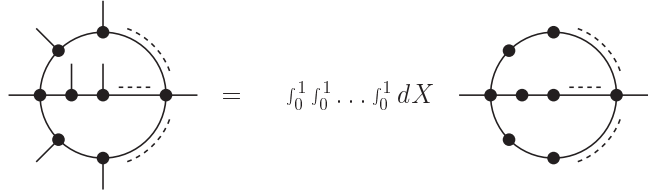}
\end{center}
\caption{Expressing generic massive two-loop Feynman diagrams as integrals
         over sunset-type functions.}
\label{fig:intosunset}
\end{figure}

In the following step, these tensor integrals are decomposed into
scalar integrals, whereby the Lorentz structure is constructed from the
vector $k^{\mu}$ and the metric tensor $g^{\mu\nu}$. This can be carried out 
systematically by decomposing the loop momenta $p$ an $q$ into components
parallel and orthogonal to $k^{\mu}$. For instance, the decomposition of all
tensor integrals up to rank three is the following:

\begin{eqnarray}
\frac{1}{[211]} & = &               {_1A_1}  
 ~~~~~~~~~~~~~~~~~~,~
\frac{p^{\mu}}{[211]}  =  k^{\mu} {_2A_1}   
 ~~~~~~~~~~~~~~,~
\frac{q^{\mu}}{[211]}  =  k^{\mu} {_3A_1}   
  \nonumber \\
\frac{p^{\mu} p^{\nu}}{[211]}  & = & \tau^{\mu\nu} {_4A_1} + g^{\mu\nu} {_4A_2}  
 ~,~
\frac{p^{\mu} q^{\nu}}{[211]}   =  \tau^{\mu\nu} {_5A_1} + g^{\mu\nu} {_5A_2}  
 ~,~
\frac{q^{\mu} q^{\nu}}{[211]}   =  \tau^{\mu\nu} {_6A_1} + g^{\mu\nu} {_6A_2}  
  \nonumber \\
\frac{p^{\mu} p^{\nu} p^{\lambda}}{[211]}  & = & 
               ( \tau^{\mu\nu} k^{\lambda} + \tau^{\mu\lambda} k^{\nu} 
                                     + \tau^{\nu\lambda} k^{\mu} ) {_7A_1}  
              + ( g^{\mu\nu} k^{\lambda} + g^{\mu\lambda} k^{\nu} 
                                     + g^{\nu\lambda} k^{\mu} )    {_7A_2} 
  \nonumber \\
\frac{q^{\mu} p^{\nu} p^{\lambda}}{[211]}  & = & 
               ( \tau^{\mu\nu} k^{\lambda} + \tau^{\mu\lambda} k^{\nu} 
                                     + \tau^{\nu\lambda} k^{\mu} ) {_8A_1} 
              + ( g^{\mu\nu} k^{\lambda} + g^{\mu\lambda} k^{\nu} 
                                     + g^{\nu\lambda} k^{\mu} )    {_8A_2}
  \nonumber \\
\frac{p^{\mu} q^{\nu} q^{\lambda}}{[211]}  & = & 
               ( \tau^{\mu\nu} k^{\lambda} + \tau^{\mu\lambda} k^{\nu} 
                                     + \tau^{\nu\lambda} k^{\mu} ) {_9A_1} 
              + ( g^{\mu\nu} k^{\lambda} + g^{\mu\lambda} k^{\nu} 
                                     + g^{\nu\lambda} k^{\mu} )    {_9A_2} 
  \nonumber \\
  & &       
 \;\;\;\;\;\;\;\;\;\;\;\;\;\;\;\;\;\;\;\;\;\;\;\;\;
 \;\;\;\;\;\;\;\;\;\;\;\;\;\;\;\;\;\;
             + ( g^{\mu\nu} k^{\lambda} + g^{\mu\lambda} k^{\nu} 
                                  - 2 g^{\nu\lambda} k^{\mu} )    {_9A_3} 
  \nonumber \\
\frac{q^{\mu} q^{\nu} q^{\lambda}}{[211]}  & = & 
               ( \tau^{\mu\nu} k^{\lambda} + \tau^{\mu\lambda} k^{\nu} 
                                     + \tau^{\nu\lambda} k^{\mu} ) {_{10}A_1} 
              + ( g^{\mu\nu} k^{\lambda}  + g^{\mu\lambda} k^{\nu}  
                                     + g^{\nu\lambda} k^{\mu} )    {_{10}A_2} 
\end{eqnarray}
In the formulae above, a loop integration $\int d^n p \; d^n q$ 
is understood, and we used the following notations:

\begin{equation}
[211] = [(p+k)^2+m_1^2]^2 \, (q^2+m_2^2) \, (r^2+m_3^2)
 ~~~,~~~
  \tau^{\mu\nu} = g^{\mu\nu} - \frac{k^{\mu}k^{\nu}}{k^2}
\end{equation}

In ref. \cite{2loopgeneral} we have shown that 
all the scalar coefficients $_iA_j$ 
involved in this tensor decomposition are directly expressible in terms 
of the ten basic functions $h_i$, up to trivial one-loop tadpole integrals.
In all the formulae above, we have considered only integrals with a special
combination of propagator powers, namely $[211]$ of eq. 6. 
Where integrals with higher powers are needed, they can be obtained 
directly by mass diferentiation. The only two-loop combination with 
lower power, $[111]$, can be obtained from $[211]$ by a recursion formula 
obtained by partial integration \cite{2loopgeneral}.

After performing these steps, the Feynman graph is decomposed into scalar
integrals expressed essentially as multiple integrals over $h_i$ functions, 
plus trivial one-loop tadpole-type contributions.
All necessary formulae to perform this reduction are given in ref. 
\cite{2loopgeneral}.
They were encoded into computer algebra programs for automatizing the 
reduction.

Once this standard integral representation is obtained for all Feynman graphs 
involved in a physical process, the final step consists in a numerical 
multi-dimensional integration of these expressions. The numerical integration
uses an adaptative deterministic algorithm, similar to the numerical 
integration for the $h_i$ functions. This ensures an efficient and precise 
evaluation of the integrals.

Several physical calculations have been performed so far by using this method.
As a two-point example to test the reduction algorithm and the reliability of 
the numerical integration, in figure 3 we show the mixed electroweak-QCD
Feynman graphs which contribute to the top quark self-energy at two-loop. 
By calculating the self-energy function
$\Sigma(p \cdot \gamma)=\Sigma_1(p \cdot \gamma) 
+ \gamma_5 \cdot \Sigma_{\gamma_5}(p \cdot \gamma)$ at two-loop,
from its imaginary part one can extract the top decay width up
to $O(\alpha_s)$, as $\Gamma_t=2 \cdot Im \Sigma_1(p \cdot \gamma = m_t)$.
Since this correction is known in an analytic form, this provides a good 
check on our two-loop algorithm. The results for the correction to the width,
obtained from the imaginary part of the two-loop self-energy, 
are given in table 1. They agree with the existing analytic results.

\begin{figure}[t]
\hspace{1.cm}
    \epsfxsize = 9cm
\begin{center}
    \epsffile{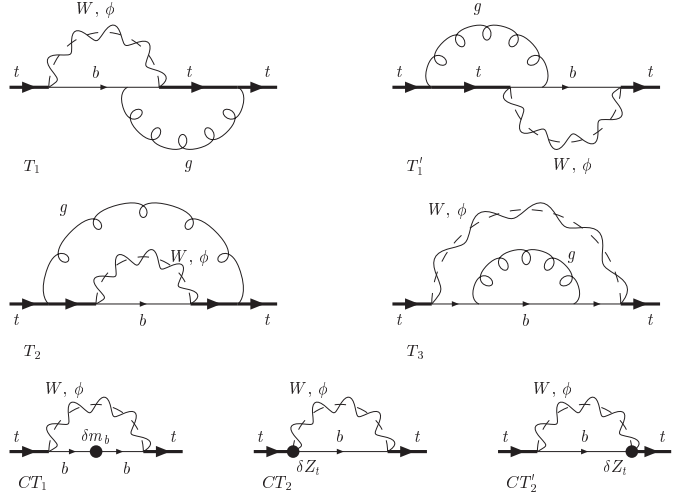}
\end{center}
\caption{{\em The two-loop Feynman graphs which
              contribute to the $b$-mass dependent correction 
              of ${\cal O}(\alpha_s g^2)$ to the top self-energy.
              Only the counterterm diagrams are shown which are needed
              for subtracting the infinities of the imaginary part of the
              self-energy, which gives the ${\cal O}(\alpha_s)$ correction
              to the $t \rightarrow W+b$ decay.}}
\end{figure}

\begin{table}
\begin{tabular}{||c||c|c|c|c|c||}                                         \hline\hline
 $m_t$ [GeV]                                      &  160  &  165  &  170  & 175   & 180    \\ \hline\hline
 $\Gamma_t^{tree}$ [GeV]           & 1.127 & 1.260 & 1.402 & 1.553 & 1.712  \\ \hline
 $\delta \Gamma_t^{1-loop}$ [GeV]  & -.092 & -.104 & -.117 & -.132 & -.149  \\ \hline\hline
\end{tabular}
\caption{The ${\ O}(\alpha_s)$ correction to the top decay  
$t \rightarrow W+b$ as obtained from the imaginary part of
the two-loop top self-energy of figure 1, integrated numerically. 
We took $G_F=1.16637 \cdot 10^{-5}$ GeV$^{-2}$, 
$m_W=80.41$ GeV, $m_b=4.7$ GeV, and $\alpha_s(m_t)=.108$.}
\end{table}

As a three-point example, in figure 4 we show the diagrams which contribute
to the top-dependent decay process $Z\rightarrow b \bar b$. 

\begin{figure}[t]
    \epsfxsize = 9cm
\begin{center}
    \epsffile{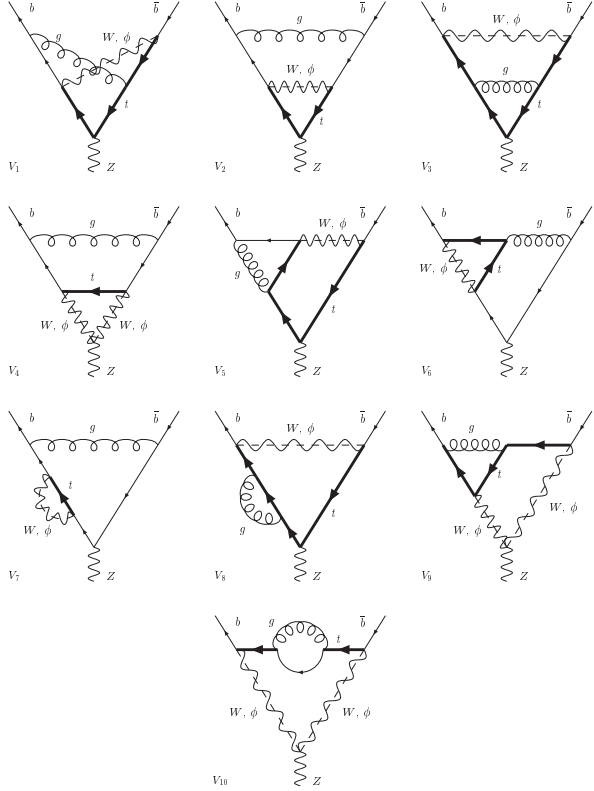}
\end{center}
\caption{Two-loop three-point diagrams contributing to 
            $Z \rightarrow b\bar b$ at ${\cal O}(\alpha_s g^2)$.}
\label{fig:diagrams}
\end{figure}

A point which 
deserves special attention in this case is the presence of IR divergences.
Because the general formulae of the $h_i$ functions are derived for general,
finite masses, IR divergences in our method are not automatically extracted 
as poles in the space-time regulator, as is costumary in QCD calculations.
IR divergences in our approach usually appear as end-point singularities in the
Feynman parameter integration over $h_i$ functions, and therefore 
require a special treatment. 

For both cases at hand -- the top quark 
self-energy and the $Z\rightarrow b \bar b$ decay -- one possible 
approach is to use a mass regulator for the gluon. While in general this 
is not possible for non-abelian theories becaus it does not preserve the 
Slavnov-Taylor identities, in the particular case of these mixed 
electroweak-QCD corrections this is a correct procedure because at 
this order the IR structure is the same as in the abelian case.

Another approach is to extract the IR structure of the graphs analytically
before numerical integration. This can be done in the form of one-loop
integrals which can be handled separately in an analytical way, by usual 
dimensional regularization. This is illustrated in fig. 5. Once the IR
divergences are extracted in the form of one-loop integrals, the two-loop
integration can be carried out numerically.

\begin{figure}
\hspace{1.cm}
    \epsfxsize = 9cm
\begin{center}
    \epsffile{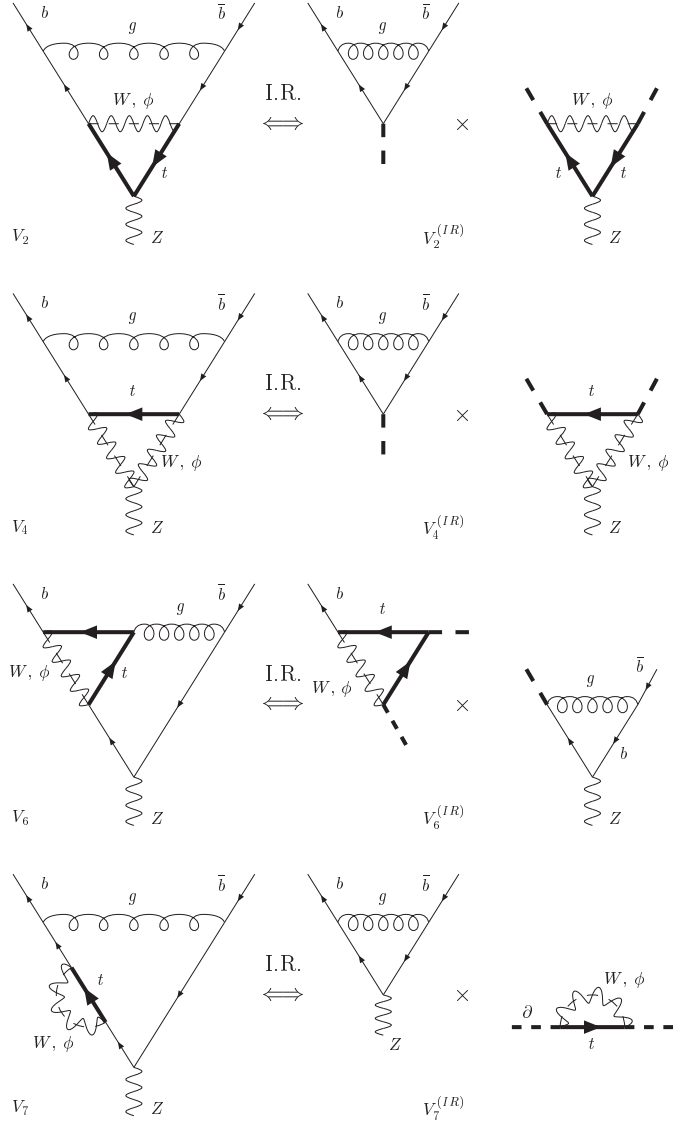}
\end{center}
\caption{{\em Extracting the infrared divergent pieces of the two-loop
              diagrams analytically. The infrared divergency of the two-loop
              diagram is the same as the infrared divergency of the 
              product of the two one-loop diagrams
              obtained by ``freezing'' the common line in the loop momenta 
              integration.}}
\label{fig:IR}
\end{figure}

We give in table 2 numerical results for all two-loop Feynman graphs
involved in this process. The numerical results are after the extraction of
the UV poles. The IR singularities are subtracted as shown in figure 5.
The numerical integration accuracy is  $10^{-3}$. The evaluation of a 
total of 78 Feynman graph evaluations with this precision
required 100 hours computing time on a 600 MHz Pentium machine.

\begin{table} 
\begin{tabular}{||l||c|c|c||}    \hline\hline
diagram           & $m_t=165$ GeV                       & $m_t=175$ GeV                       & $m_t=185$ GeV                       \\ \hline\hline
$V_1$             & $ -1.009            \cdot 10^{-3}$  & $-7.187             \cdot 10^{-4}$  & $ -4.057            \cdot 10^{-4}$  \\
$V_2-V_2^{(IR)}$  & $(-2.873 + i 2.122) \cdot 10^{-3}$  & $(-2.490 + i 1.147) \cdot 10^{-3}$  & $(-2.087 + i .09274)\cdot 10^{-3}$  \\
$V_3$             & $  1.545            \cdot 10^{-3}$  & $ 2.255             \cdot 10^{-3}$  & $  3.034            \cdot 10^{-3}$  \\
$V_4-V_4^{(IR)}$  & $( 1.215 - i 2.481) \cdot 10^{-2}$  & $( 1.242 - i 2.570) \cdot 10^{-2}$  & $( 1.266 - i 2.660) \cdot 10^{-2}$  \\
$V_5$             & $  2.107            \cdot 10^{-2}$  & $ 2.469             \cdot 10^{-2}$  & $  2.861            \cdot 10^{-2}$  \\
$V_6-V_6^{(IR)}$  & $( 3.089 - i 4.257) \cdot 10^{-2}$  & $( 3.500 - i 4.824) \cdot 10^{-2}$  & $( 3.950 - i 5.445) \cdot 10^{-2}$  \\
$V_7-V_7^{(IR)}$  & $(-.7778 + i 1.281) \cdot 10^{-2}$  & $(-.8001 + i 1.349) \cdot 10^{-2}$  & $(-.8232 + i 1.420) \cdot 10^{-2}$  \\
$V_8$             & $ -1.059            \cdot 10^{-3}$  & $-1.474             \cdot 10^{-3}$  & $ -1.942            \cdot 10^{-3}$  \\
$V_9$             & $  6.289            \cdot 10^{-2}$  & $ 6.703             \cdot 10^{-2}$  & $  7.143            \cdot 10^{-2}$  \\
$V_{10}$          & $ -1.402            \cdot 10^{-2}$  & $-1.389             \cdot 10^{-2}$  & $ -1.380            \cdot 10^{-2}$  \\ \hline\hline
\end{tabular} 
\caption{Numerical values for the two-loop diagrams shown in figure~\ref{fig:diagrams}. 
$V_1$--$V_{10}$ are the sums of the corresponding $W$ and $\phi$ 
exchange graphs.
An overall color and coupling constant factor of 
$i \gamma_{\mu} (1-\gamma_5) \alpha_s (g^3/12 \cos \theta_W)$ is understood.
The UV and IR divergences are removed  
as discussed in the text.}
\end{table}

To conclude, we developed an algorithm for the tensor reduction of 
massive two-loop diagrams. It applies in principle to any massive 
two-loop graph, and it can be automatized in the form of a computer 
algebra program. The tensor decomposition algorithm results in a 
multi-dimensional integral over a set of ten basic functions $h_i$, 
which are defined in terms of one-dimensional integral representations.
We described the numerical methods which we used for carrying out 
the remaining integrations. 

We have shown how these techniques work in the case of two realistic
calculations of mixed electroweak-QCD radiative corrections. The first example 
is the two-loop top quark self-energy from which the $O(\alpha_s)$ correction
to the top quark decay width can be extracted and compared with the analytical
result. The second example is a three-point calculation involving all two-loop
diagrams which contribute to the top-dependent decay process 
$Z\rightarrow b \bar b$.
Thus we have shown that the techniques we described can be used in realistic
calculations, where several internal mass and external momenta scales 
are involved. 

This approach works for any such combination of kinematic 
variables, apart from possible infrared complications.
In the context of the $Z \rightarrow b \bar b$ example, 
we discussed the analytical separation of the infrared divergencies.
Within our two-loop methods, if a process involves infrared singularities,
these have to be dealt with in a special way because the numerical
nature of our methods.

\end{document}